\documentclass[runningheads]{llncs}
\usepackage[T1]{fontenc}
\usepackage{graphicx}
\usepackage{pifont}

\usepackage{amsmath,amsfonts,bm}

\def\eqref#1{equation~\ref{#1}}
\def\1{\bm{1}}

\DeclareMathAlphabet{\mathsfit}{\encodingdefault}{\sfdefault}{m}{sl}
\SetMathAlphabet{\mathsfit}{bold}{\encodingdefault}{\sfdefault}{bx}{n}

\usepackage[framemethod=TikZ]{mdframed}
\mdfdefinestyle{MyFrame}{%
    linecolor=black,
    outerlinewidth=0.3pt,
    roundcorner=5pt,
    skipabove = 5.5pt,
    skipbelow = 5.5pt,
    innertopmargin=5.5pt, %\baselineskip,
    innerbottommargin=5.5pt, %\baselineskip,\baselineskip,
    innerrightmargin=5.5pt,
    innerleftmargin=5.5pt,
    backgroundcolor=bg, %gray!25!white
}

\newcommand{\name}{\textsc{ModelWisdom}}

\usepackage{hyperref}
\usepackage{url}
\usepackage{graphicx}

\usepackage{tcolorbox}
\usepackage{listings}
\usepackage{xcolor,colortbl}

\usepackage{xspace}

\usepackage{enumitem}
\setlist[itemize]{leftmargin=1em}

\usepackage{caption}
\usepackage{subcaption}

\begin{document}
\definecolor{bg}{HTML}{F8F9FB}  % FAFAFA

\title{\name: 
An Integrated Toolkit for TLA$^{+}$ Model Visualization, Digest and Repair
% An LLM-Enhanced Interpretable Model Checking Toolkit
}

\titlerunning{\name}
% If the paper title is too long for the running head, you can set
% an abbreviated paper title here
%
\author{Zhiyong Chen\thanks{These authors contributed equally to this work.}\inst{1}\orcidID{0009-0005-2555-5728} \and
Jialun Cao*\inst{2,3}\orcidID{0000-0003-4892-6294} \and
Chang Xu\thanks{Corresponding author.}\inst{1}\orcidID{0000-0002-6299-4704} \and
Shing-Chi Cheung\inst{2,3}\orcidID{0000-0002-3508-7172}}

\authorrunning{\name}
% First names are abbreviated in the running head.
% If there are more than two authors, 'et al.' is used.
%
\institute{State Key Laboratory for Novel Software Technology, \\ Nanjing University, Nanjing, China \\
\email{zhiyongchen@smail.nju.edu.cn, changxu@nju.edu.cn} \and
Department of Computer Science and Engineering, The Hong Kong University of Science and Technology, Hong Kong, China \and Guangzhou HKUST Fok Ying Tung Research Institute, Guangzhou, China
\email{jialuncao@ust.hk, scc@cse.ust.hk}}
\maketitle              % typeset the header of the contribution
\begin{abstract}
 % ModelWisdom is an modern integrated toolkit designed to enhance the usability and accessibility of TLA+ modeling through three core modules: ModelVisualizer, ModelRepair, and ModelDigest. ModelVis provides intuitive visualization of TLA+ models, enabling users to explore system behaviors graphically. ModelRepair and ModelDigest leverages large language models to assists in automatically fixing inconsistencies or errors within models, and to generate natural language explanations of model behaviors. Together, these components empower system designers and verifiers to better understand and communicate TLA+ specifications, thereby advancing the practical adoption of formal methods in system design.

Model checking in TLA+ provides strong correctness guarantees, yet practitioners continue to face significant challenges in interpreting counterexamples, understanding large state-transition graphs, and repairing faulty models. These difficulties stem from the limited explainability of raw model-checker output and the substantial manual effort required to trace violations back to source specifications. Although the TLA+ Toolbox includes a state diagram viewer, it offers only a static, fully expanded graph without folding, color highlighting, or semantic explanations, which limits its scalability and interpretability. We present \name, an interactive environment that uses visualization and large language models to make TLA+ model checking more interpretable and actionable. \name~offers: (i) Model Visualization, with colorized violation highlighting, click-through links from transitions to TLA+ code, and mapping between violating states and broken properties; (ii) Graph Optimization, including tree-based structuring and node/edge folding to manage large models; (iii) Model Digest, which summarizes and explains subgraphs via large language models (LLMs) and performs preprocessing and partial explanations; and (iv) Model Repair, which extracts error information and supports iterative debugging. Together, these capabilities turn raw model-checker output into an interactive, explainable workflow, improving understanding and reducing debugging effort for nontrivial TLA+ specifications. The website to \name~ is available: \url{https://model-wisdom.pages.dev}. A demonstrative video can be found at \url{https://www.youtube.com/watch?v=plyZo30VShA}. 

\end{abstract}

\section{Introduction}\label{sec:intro}
Model checking~\cite{baier2008principles,clarke1997model} in TLA+~\cite{lamport2002specifying,merz2007specification,lamport1999specifying} provides strong correctness guarantees, yet practitioners often struggle to interpret the large state-transition graphs and counterexamples produced by the model checker. While the TLA+ Toolbox~\cite{kuppe2019tlaplustool} includes a built-in state diagram viewer, it displays graphs in a static, fully expanded form without node or edge folding, colorized violation cues, or semantic explanations. As a result, users need to sift through dense graphs manually, trace transitions back to the code, and connect violation states to the temporal properties they violate -- tasks that become increasingly difficult as models grow in size and complexity. 

\begin{figure}[h!]
\begin{center}
%\framebox[4.0in]{$\;$}
\includegraphics[width=1\linewidth]{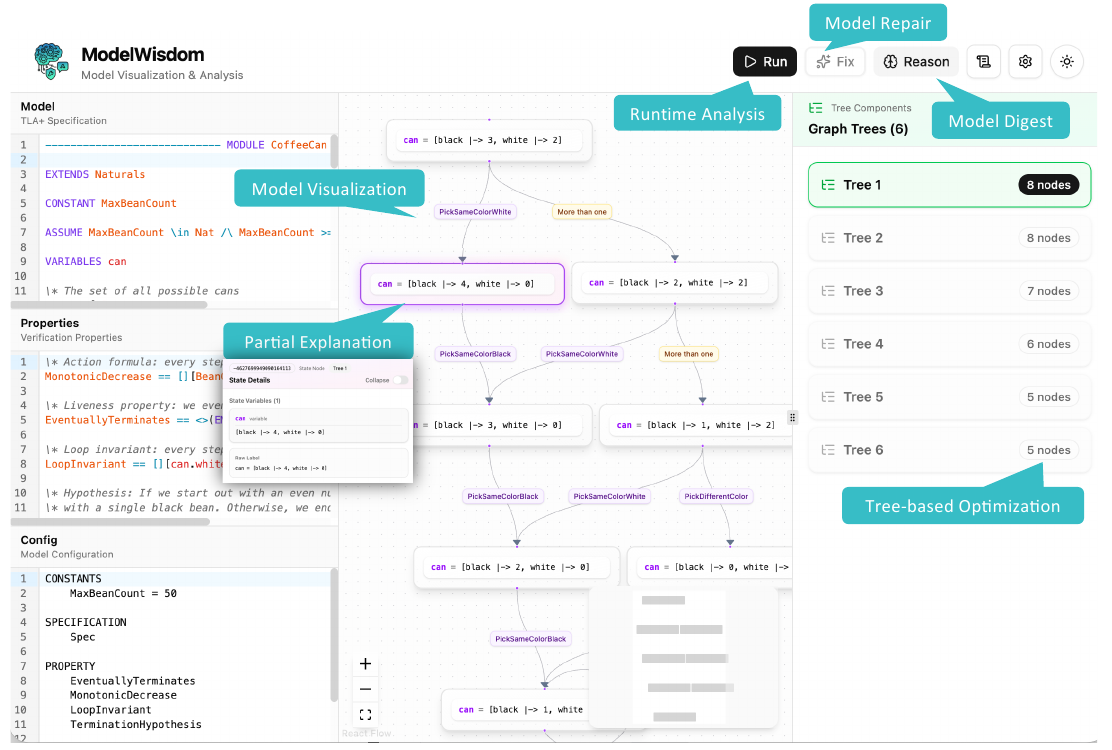}
\end{center}
\setlength{\abovecaptionskip}{-0pt}
\caption{Interface of \name with Highlight Features}
\label{fig:interface}
\end{figure}

These challenges echo long-standing usability problems in model checking tools~\cite{havelund2001monitoring,holzmann1997model}, where understanding diagnostic output often becomes harder than writing the specification itself. Meanwhile, recent advances in large language models (LLMs) open new opportunities to turn these verification artifacts into interactive, interpretable representations. While prior work has used LLMs for specification synthesis~\cite{wen2024enchanting,ma2025specgen,kogler2024reliable,zhao2024nl2ctl,alhanahnah2025empirical,cao-etal-2025-informal} or proof assistance~\cite{yang2023leandojo,song23towards,qu2024recursive}, little attention has been given to enhancing the visibility, navigability, and explainability of model checking. Complementing algorithmic improvements, improving the human–tool interface is increasingly recognized as a critical step toward broader industry adoption of formal verification.

To address these gaps, we present \name, an LLM-assisted interactive environment that enhances the interpretability, explorability, and repairability of TLA+ model-checking results. \name~integrates visualization, structural abstraction, automated debugging support, and model summarization into a unified workflow. Its contributions are fourfold:

\ding{52} \textbf{Model Visualization \& Explainability}: Colorized violation nodes, click-to-navigate edges linking to source code, and direct mapping from violating states to broken properties.

\ding{52} \textbf{Performance-Oriented Optimization}: Scalable exploration through tree-structured graph layouts, node and edge folding, and automated graph compaction, as well as support for identifying homogeneous node clusters that can guide parameterized verification~\cite{zuck2004model,john2013parameterized,arons2001parameterized,chen2006general,chou2004simple}.

\ding{52} \textbf{Model Digest}: Graph summarization, preprocessing, and selective subgraph explanation using LLM-guided analysis.

\ding{52} \textbf{Model Repair Assistance}: Structured extraction of error information, support for both single-pass and multi-pass repair strategies, and an interactive history panel that records intermediate fixes and enables systematic, traceable debugging of faulty specifications.

\section{\name}\label{sec:modelwisdom}

\name~is deployed on the website (\url{https://model-wisdom.pages.dev}), comprising approximately 7,000 lines of fully typed TypeScript and Python code. It consists of two main components: the primary web application and a server responsible for executing the TLC (i.e., TLA+ Checker). The supporting LLMs include GPT and Claude families (e.g., gpt-4 and Claude 3.7 Sonnet), and can be set on the website. 

\subsection{Model Visualizer}

% Feature
% 1. Colorize violation
% 2. 点击边，关联代码
% 3. 点击 violation 节点，关联 property

% Performance Optimization (usb
% 1. 分树显示
% 2. 节点折叠
% 3. 边折叠

% parameterized verification

In this paper, we introduce ModelVisualizer, a tool designed to enhance both the visualization and performance of state transition graph analysis in the context of TLA+ verification. ModelVisualizer enables users to input a TLA+ model, specify properties, and configure parameters for verification via TLC. Upon completion, ModelVisualizer automatically collects the resulting state transition graph and tightly integrates it with the underlying TLA+ code. 

One of our insights is to bridge model code and interactive exploration: ModelVisualizer highlights violation nodes with distinct colors, provides click-through navigation from transitions directly to the corresponding lines of TLA+ code, and offers a direct mapping between violating states and the specific properties they break. This approach allows users to quickly trace errors from state graph nodes to concrete model definitions. Similar visualization features are also supported by the TLA+ Toolkit~\cite{kuppe2019tlaplustool} and UPPAAL~\cite{larsen1997uppaal,behrmann2004tutorial,behrmann2006uppaal,behrmann2006tutorial}, while \name was empowered by LLMs, offering explaining and repairing features. 

To address the challenge of visualizing large and complex state graphs without sacrificing performance, ModelVisualizer incorporates several graph optimization techniques. These include tree-based structuring for hierarchical clarity, node and edge folding to reduce visual clutter, and graph compaction for efficient rendering. The core idea behind these optimization strategies is lazy rendering, where only the currently interacted portions of the graph are generated and displayed on demand.

\begin{figure}[t!]
\begin{center}
%\framebox[4.0in]{$\;$}
\includegraphics[width=0.8\linewidth]{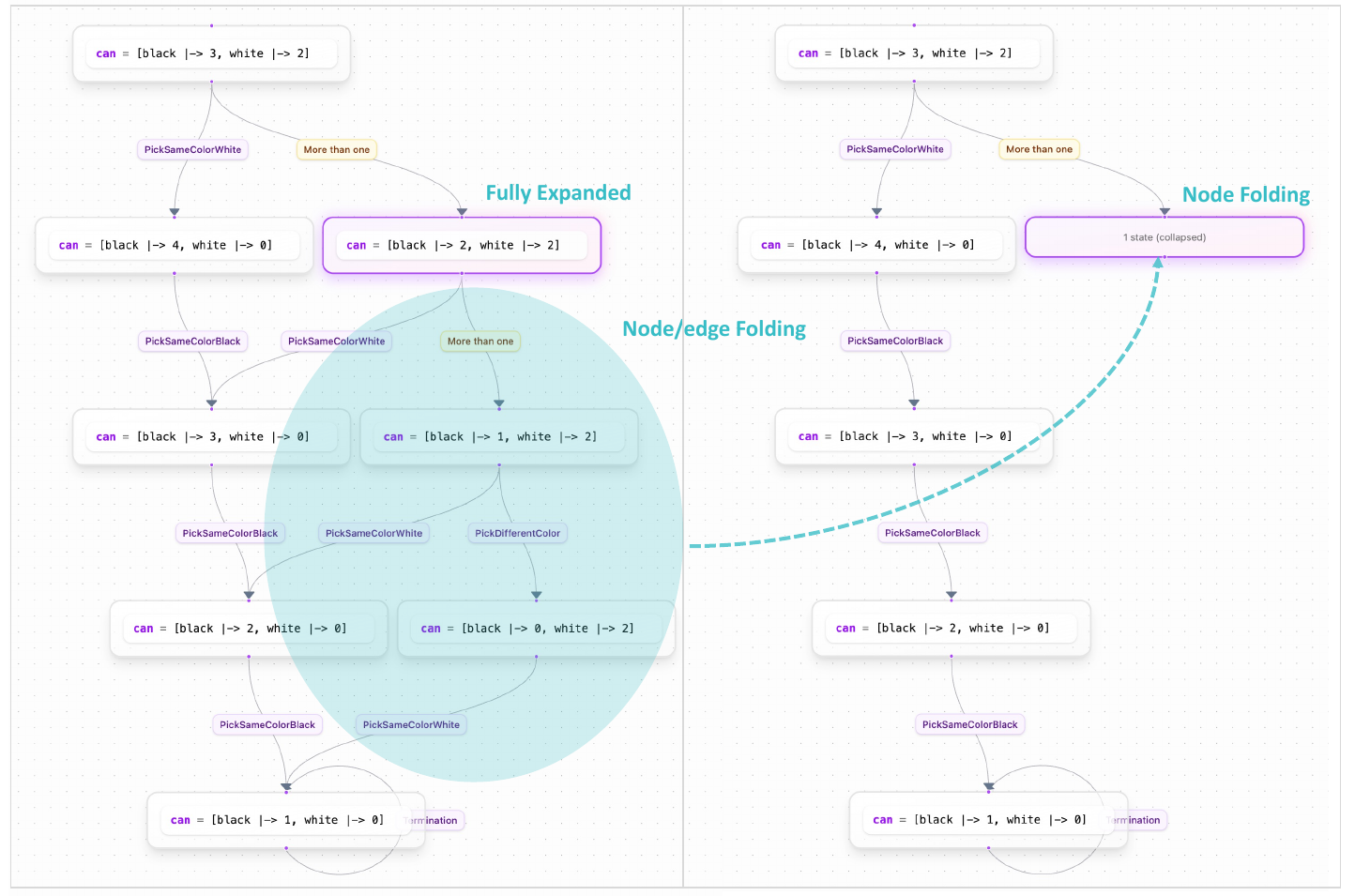}
\end{center}
\setlength{\abovecaptionskip}{-0pt}
\caption{Interface of \name with node folding}
\label{fig:node-folding}
\end{figure}

\begin{figure}[t!]
\centering

%\framebox[4.0in]{$\;$}
\begin{subfigure}[b]{1\textwidth}
     \includegraphics[width=\textwidth, trim=0 160pt 0 160pt, clip]{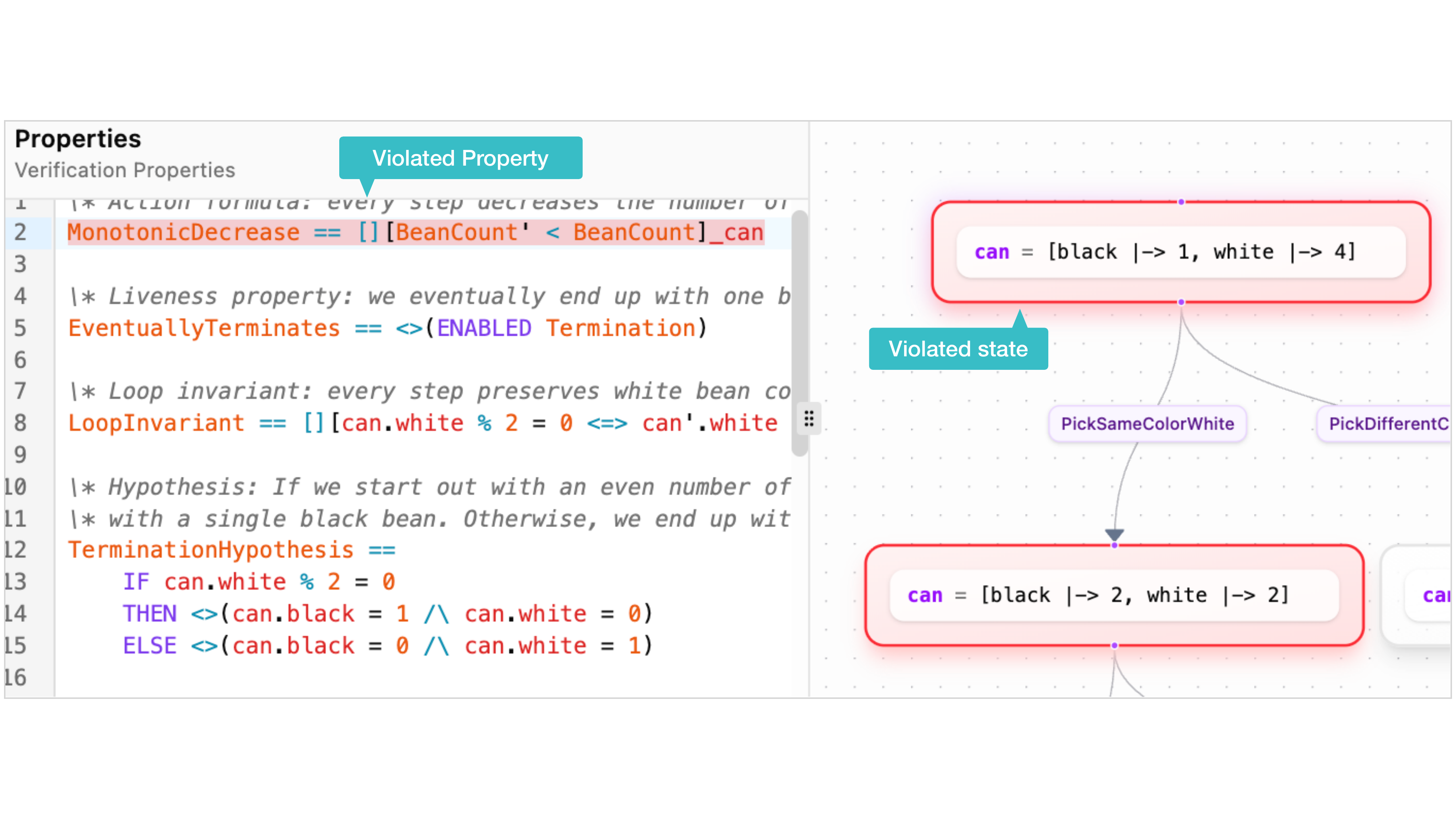}
     \caption{Violated Property Visualization}
     \label{fig:violated-property}
\end{subfigure}
\begin{subfigure}[b]{0.49\textwidth}
     \includegraphics[width=\textwidth, trim=105pt 0 105pt 0, clip]{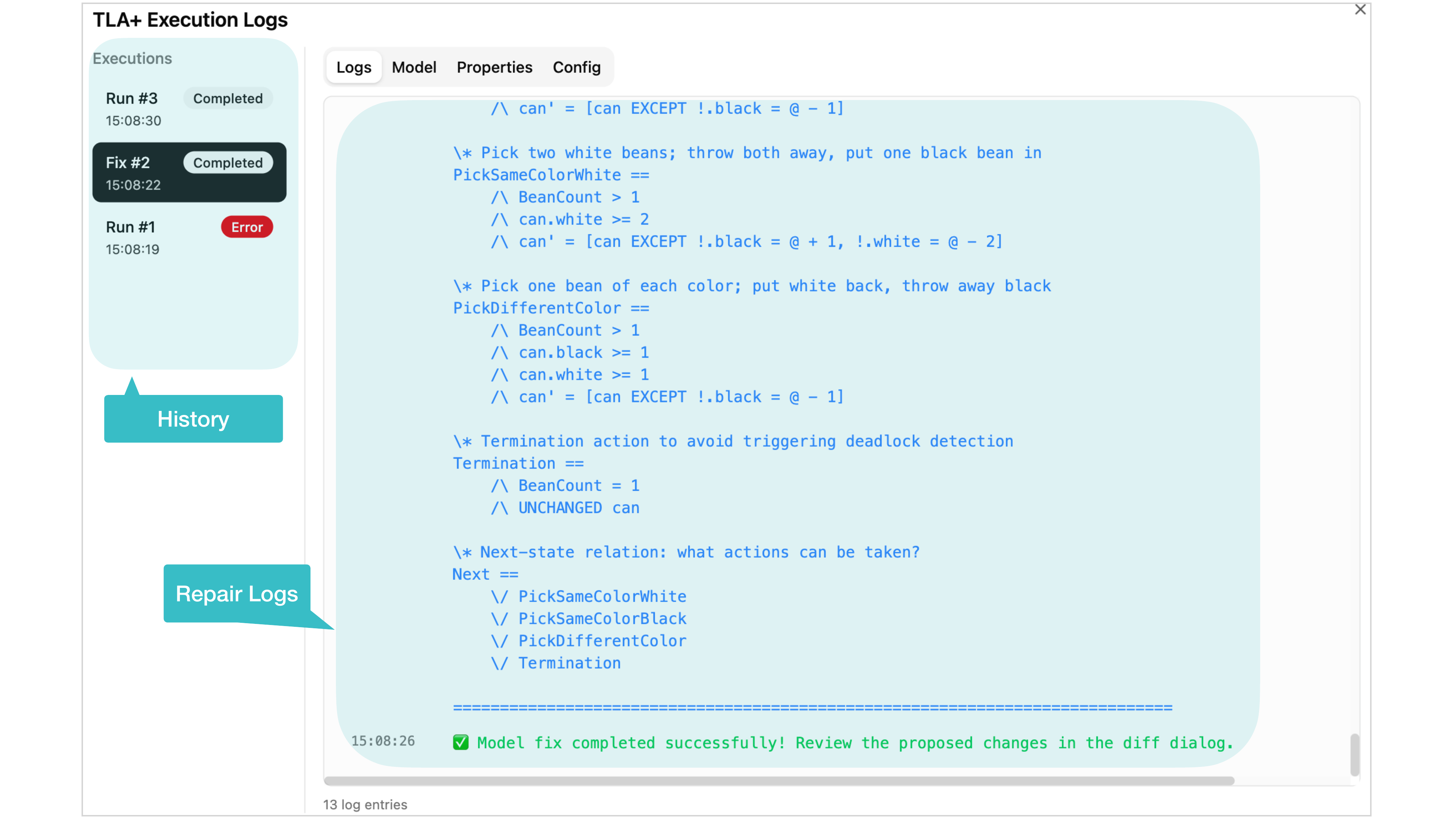}
     \caption{ModelRepair Dialog}
     \label{fig:model-repair}
\end{subfigure}
\begin{subfigure}[b]{0.49\textwidth}
     \includegraphics[width=\textwidth, trim=105pt 0 105pt 0, clip]{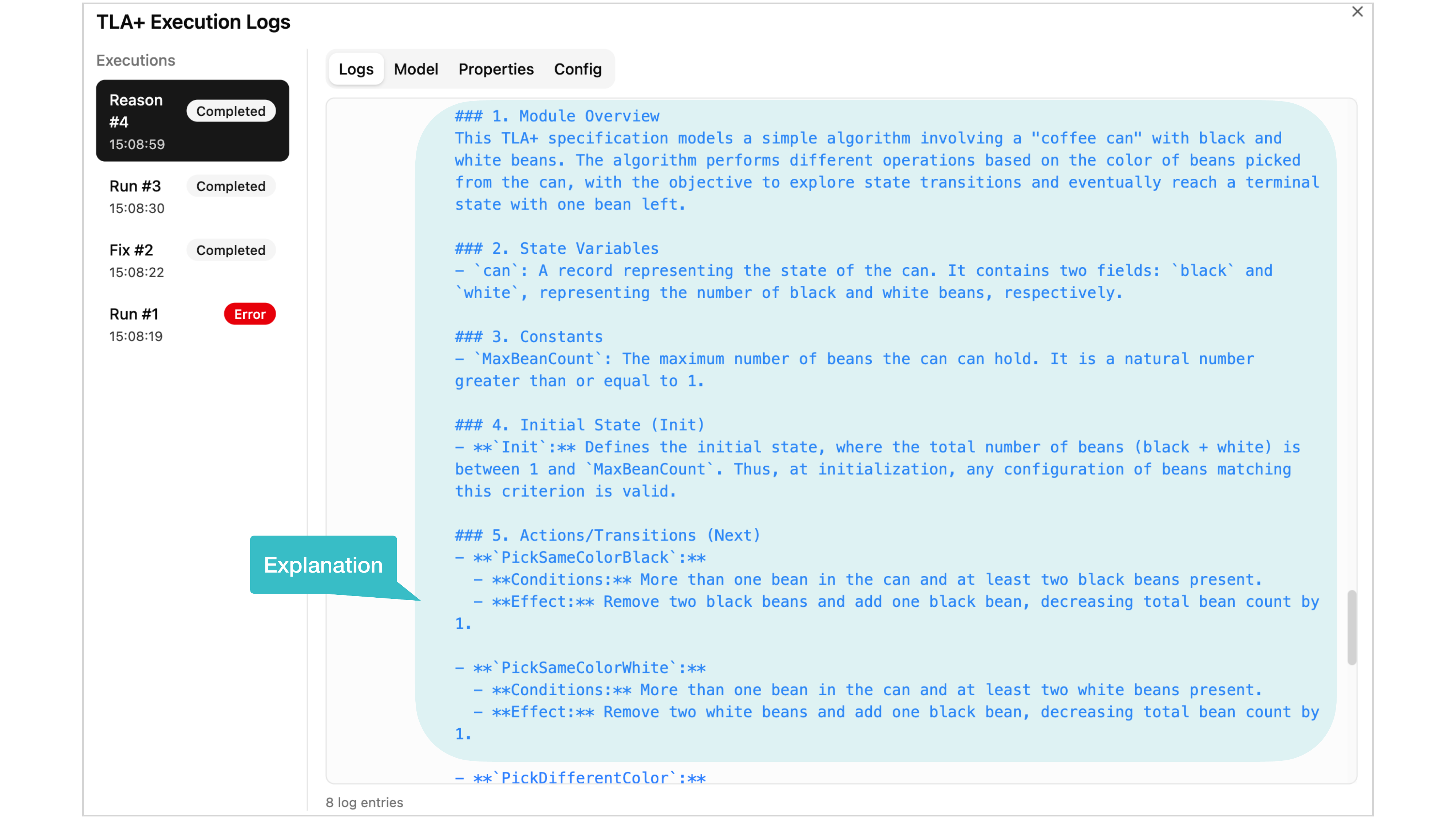}
     \caption{ModelDigest Dialog}
     \label{fig:model-digest}
\end{subfigure}

\caption{Interface of \name with highlighting violated properties, ModelRepair, and ModelDigest}
\end{figure}

\textit{Demonstration 1.} Figure \ref{fig:interface} demonstrates the interface of \name~ under the TLA+ official example of \texttt{CoffeeCan}\footnote{\url{https://github.com/tlaplus/Examples/blob/master/specifications/CoffeeCan/CoffeeCan.tla}}. The interactive state graph is visualized in the middle of the application after successful runtime analysis. Meanwhile, the ModelVisualizer only renders one tree at a time for performance, and the users can toggle the trees on the right-handle panel. Figure \ref{fig:node-folding} demonstrates the node-folding technique, which can be utilized to track state transitions incrementally, or to collapse irrelevant paths for enhanced rendering efficiency. Figure \ref{fig:violated-property} demonstrates the highlight of violated states and properties, where we deliberately modify the postcondition statement of the \texttt{PickSameColorWhite} action from \texttt{can' = [can EXCEPT !.black = @ + 1, !.white = @ - 2]} to \texttt{can' = [can EXCEPT !.black = @ + 1, !.white = @ - 1]}.

\subsection{Model Digest}

% 1. 图总结
% 2. Preprocess
% 3. 部分选择解释

In this paper, we offer ModelDigest to explain TLA+ models as a whole or in part. ModelDigest enables users to highlight specific code segments of interest directly within the editor, facilitating targeted explanations. Prior to engaging LLMs, ModelDigest automatically extracts key structural information from the state graph, including initial and terminal states, cycles, and the most frequent transitions—to provide essential context. Leveraging this information, LLMs generate comprehensive explanations of the TLA+ model, covering aspects including the overall model structure, state variables, constants, actions, transitions, and invariants.

\textit{Demonstration 2.} Figure \ref{fig:model-digest} demonstrates the explanation of ModelDigest for the repaired model in Figure \ref{fig:violated-property}. ModelDigest provides LLMs with the complete TLA+ specification as well as a summary of the state graph, including the initial state (\texttt{can = [black |-> 0, white |-> 5]}) and the terminal state (\texttt{can = [black |-> 0, white |-> 1]}). Finally, ModelDigest's explanation includes the semantics of each action in natural language. For example, for the action \texttt{PickSameColorBlack}, the explanation is as follows: (1) Removes two black beans and adds one black bean back; (2) Preconditions: more than one bean in total and at least two black beans; (3) Postcondition: one black bean is removed.

\subsection{Model Repair}

% 1. 错误信息提取 (error detection)
% 2. 小循环 (fix count), single-pass
% 3. history of repair, multiple-pass

In this paper, we propose ModelRepair. It is designed to facilitate the debugging and iterative repair of TLA+ models, inspired by the TLA+ debugger~\cite{kuppe2022tla}. ModelRepair automatically extracts error messages generated during TLC checking and leverages LLMs to repair the TLA+ model when syntax errors or violations of properties are detected. The tool supports both single-pass and multi-pass repair workflows. In the single-pass repair mode, ModelRepair provides an interactive, one-shot repair experience, where the user can review and accept the suggested fix for a given error. In contrast, the multi-pass repair workflow enables iterative debugging by automatically engaging in multiple rounds of interaction between LLMs and TLC. During this process, ModelRepair maintains a history panel that records each repair attempt, allowing users to track changes and understand the evolution of the model. The iterative process continues until the model is free of errors or a predefined repair limit is reached.

\textit{Demonstration 3.} Figure \ref{fig:model-repair} illustrates the repair and execution history of ModelRepair in the case of the invalid models in Figure \ref{fig:violated-property}. ModelRepair provides the LLMs with comprehensive information regarding the incorrect specification as well as the extracted error message. Subsequently, ModelRepair successfully corrects the erroneous statement \texttt{!.white = @ - 1}, reverting it to the correct one \texttt{!.white = @ - 2}. On the right side, detailed logs are provided, including the specific content of the LLM-driven dialog as well as the repaired model.

\section{Future Work}\label{sec:future-work}

In future work, we plan to extend \name~ beyond TLA+ to support a broader class of model checking languages and verification frameworks. Many of the challenges addressed by \name, such as interpreting counterexamples, navigating large state-transition graphs, summarizing behaviors, and assisting with repair, are shared across specification languages like Promela, Alloy, NuSMV, and UPPAAL and other state-transition models. By generalizing our visualization pipeline, LLM-assisted summarization modules, and repair interfaces, we aim to create a unified, language-agnostic environment that brings explainability and interactive assistance to diverse verification ecosystems. This expansion will not only increase the tool’s applicability but also deepen our understanding of how LLMs and visualization techniques can systematically enhance the broader practice of formal verification.

\section{Conclusion}\label{sec:conclusion}

\name~ enhances the usability and interpretability of TLA+ model checking by integrating scalable graph visualization, semantic summarization, and automated repair assistance. By combining interactive state-space exploration with LLM-driven explanations, the tool lowers the cognitive barrier to understanding counterexamples and accelerates the process of correcting faulty specifications. These capabilities go beyond the existing TLA+ Toolbox, offering colorized violation highlighting, collapsible graph structures, and property-aware diagnostics. Together, these features make \name~ a practical and powerful companion for engineers and researchers working with complex TLA+ specifications.

%
% ---- Bibliography ----
%
% BibTeX users should specify bibliography style 'splncs04'.
% References will then be sorted and formatted in the correct style.
%
\bibliographystyle{splncs04}
\bibliography{fm2026_conference}

\end{document}